\documentclass[letter]{aa} 

\usepackage{graphicx}
\usepackage{txfonts}
\usepackage[]{hyperref}
\begin{document}

   \title{Milliarcsecond-scale radio structure of the most distant BL Lac object candidate at redshift 6.57}


   \author{S.~Frey
          \inst{1,2,3}
          \and
          Y.~Zhang\inst{4,5}
          \and
          K.~Perger\inst{1,2}
          \and
          T.~An\inst{4,5}
          \and
          K.~\'E.~Gab\'anyi\inst{6,7,1,2}
          \and
          L.~I.~Gurvits\inst{8,9}
          \and
          C.-Y.~Hwang\inst{10}
          \and
          E.~Koptelova\inst{10}
          \and
          Z.~Paragi\inst{8}
          \and
          J.~Fogasy\inst{1,2}
          }

   \institute{
Konkoly Observatory, HUN-REN Research Centre for Astronomy and Earth Sciences, Konkoly Thege Mikl\'os \'ut 15-17, H-1121 Budapest, Hungary\\
\email{frey.sandor@csfk.org}
     \and
CSFK, MTA Centre of Excellence, Konkoly Thege Mikl\'os \'ut 15-17, H-1121 Budapest, Hungary
    \and
Institute of Physics and Astronomy, ELTE E\"otv\"os Lor\'and University, P\'azm\'any P\'eter s\'et\'any 1/A,
H-1117 Budapest, Hungary
     \and
Shanghai Astronomical Observatory, Chinese Academy of Sciences, 80 Nandan Road, Shanghai 200030, P.R. China
    \and 
Key Laboratory of Radio Astronomy and Technology, Chinese Academy of Sciences, A20 Datun Road, Chaoyang District, Beijing 100101, P.R. China    
    \and
Department of Astronomy, Institute of Physics and Astronomy, ELTE E\"otv\"os Lor\'and University, P\'azm\'any P\'eter s\'et\'any 1/A, H-1117 Budapest, Hungary
     \and
HUN-REN--ELTE Extragalactic Astrophysics Research Group, E\"otv\"os Lor\'and University, P\'azm\'any P\'eter s\'et\'any 1/A, H-1117 Budapest, Hungary
     \and
Joint Institute for VLBI ERIC, Oude Hoogeveensedijk 4, 7991 PD Dwingeloo, The Netherlands
     \and
Faculty of Aerospace Engineering, Delft University of Technology, Kluyverweg 1, 2629 HS Delft, The Netherlands
     \and
Graduate Institute of Astronomy, National Central University, Taoyuan City, 32001, Taiwan
             }

   \date{Received 14 November 2023 / Accepted 25 December 2023}

 
  \abstract
   {The existence of accreting supermassive black holes up to billions of solar masses at early cosmological epochs (in the context of this work, redshifts $z \gtrsim 6$) requires very fast growth rates which is challenging to explain. The presence of a relativistic jet can be a direct indication of activity and accretion status in active galactic nuclei (AGN), constraining the radiative properties of these extreme objects. However, known jetted AGN beyond $z\sim 6$ are still very rare.}
   {The radio-emitting AGN J2331$+$1129 has recently been claimed as a candidate BL Lac object at redshift $z=6.57$, based on its synchrotron-dominated emission spectrum and the lack of ultraviolet/optical emission lines. It is a promising candidate for the highest-redshift blazar known to date. The aim of the observations described here was to support or refute the blazar classification of this peculiar source.}
   {We performed high-resolution radio interferometric imaging observations of J2331$+$1129 using the Very Long Baseline Array at $1.6$ and $4.9$~GHz in 2022 Feb.}
   {The images revealed a compact but slightly resolved, flat-spectrum core feature at both frequencies, indicating that the total radio emission is produced by a compact jet and originates from within a central 10-pc scale region. While these are consistent with the radio properties of a BL Lac object, the inferred brightness temperatures are at least an order of magnitude lower than expected from a Doppler-boosted radio jet, leaving the high-redshift BL Lac identification still an open question.}
   {}


   \maketitle
%

\section{Introduction}
\label{intro}

The most powerful active galactic nuclei (AGN) can be observed from vast distances, at redshifts beyond $z=10$ \citep{2023ApJ...955L..24G,2023NatAs.tmp..223B}, corresponding to about $3.5\%$ of the present age of the Universe. These objects are powered by material accretion onto the central supermassive black holes (SMBHs) residing in their host galaxies. The existence of black holes reaching the mass of billions of solar masses at $z \sim 6-7$ poses challenges for models describing the formation and early growth of SMBHs \citep[e.g.][]{2021NatRP...3..732V}. Recent discoveries of AGN at extremely high redshifts ($z \approx 10$) favour models with massive ($\sim 10^4-10^5\,\mathrm{M}_{\odot}$) black hole seeds such as resulting from direct collapse of gravitationally unstable gas clouds \citep[e.g.][]{2006MNRAS.371.1813L}, merger of/accretion onto primordial black holes \citep[e.g.][]{2022ApJ...926..205C}, or spherically symmetric accretion into the combined potential of a stellar-mass seed black hole at the centre of a dark matter halo \citep{2023arXiv231006898S}.

According to the updated catalogue\footnote{\url{https://cdsarc.cds.unistra.fr/viz-bin/cat/J/other/FrASS/4.9}, accessed on 2023 Nov 14} of \cite{2017FrASS...4....9P}, the number of currently known $z>6$ quasars with measured spectroscopic redshift is nearly 300. From these, only 15 sources have been individually detected in radio bands, although many of the remaining objects may have some level of weak AGN-related radio emission \citep{2019MNRAS.490.2542P,2024MNRAS.527.3436P}. It is expected that radio-emitting AGN will be found up to $z\approx15$ in the coming decades, thanks to new sensitive radio and optical instruments \citep{2015aska.confE..93A,2023MNRAS.519.2060I,2024MNRAS.527L..37L}. 

In general, the relative minority, about one-tenth of quasars \citep[e.g.][]{2002AJ....124.2364I} produce relativistic plasma jets that originate from the close vicinity of the central SMBH. These are strong radio emitters via the synchrotron process of charged particles accelerated to relativistic speeds and moving in magnetic fields. Such sources are important targets for very-long-baseline interferometric (VLBI) observations that are capable of imaging the radio emission from compact jets down to pc scales -- better than any other technique in astronomy at present. Radio measurements are not affected by dust obscuration in the young host galaxies and provide a unique insight into jet launching and emission. These distant AGN are signposts of the earliest occurrence of SMBH activity in the Universe.  Studies of these objects contribute to understanding the evolution of jetted AGN, the role jets play in the co-evolution SMBHs and their host galaxies, and may even help evaluate cosmological models.

Blazars form a special class of AGN with their jets believed to be closely aligned with the line of sight and thus their emission is enhanced by Doppler boosting. About half of the currently known $21$ distant radio quasars at $z \ge 5.8$, $10$ objects have been observed, and $9$ detected with VLBI \citep{2003MNRAS.343L..20F,2005A&A...436L..13F,2008A&A...484L..39F,2011A&A...531L...5F,2008AJ....136..344M,2018ApJ...861...86M,2021AJ....161..207M,2014A&A...563A.111C,2017ApJ...835L..20W,2020A&A...643L..12S,2022A&A...662L...2Z,2022ApJ...939L...5L}. Among the latter, the only blazar candidate is PSO~J030947.49$+$271757.31 at $z=6.10$ \citep{2020A&A...643L..12S}, based on its prominent one-sided core--jet structure. The other typical $z \ga 6$ VLBI radio sources known to date have compact but somewhat resolved structure and steep radio spectrum, indicating unbeamed jet emission and suggesting young age. The double structure of the quasar NDWFS~J14276$+$3312 $(z=6.12)$ with components separated by $\sim 160$~pc \citep{2008A&A...484L..39F,2008AJ....136..344M} is reminiscent of compact symmetric objects known to be young ($\la 10^3-10^4$~yr) radio sources.

FIRST~J233153.20$+$112952.11 (J2331$+$1129 hereafter) is a source from the Faint Images of the Radio Sky at Twenty centimeters (FIRST) survey \citep{1995ApJ...450..559B} catalogue, with flux density $S_{\mathrm{1.4\,GHz}}=1.85$~mJy. This object was recently identified as a high-redshift AGN candidate by cross-matching $z_\mathrm{PS1}$-band dropouts from the Panoramic Survey Telescope and Rapid Response System 1 \citep[Pan-STARRS1,][]{2016arXiv161205560C} with FIRST radio sources \citep{2022ApJ...929L...7K}. The source has counterparts in the near- and mid-infrared, as well as in other radio surveys. Its radio spectrum is flat within a relatively narrow range of frequencies, between the observed $888$~MHz and $3$~GHz (spectral index $\alpha=-0.01 \pm 0.06$, defined here as $S \propto \nu^{\alpha}$, where $\nu$ is the frequency), with flux densities around $2$~mJy. The optical-to-radio spectral energy distribution of J2331$+$1129 is dominated by the synchrotron emission of the jet and closely resembles those of BL Lac objects. The non-thermal ultraviolet/optical continuum has a spectral index $\alpha_{\rm opt}=1.43\pm0.23$. The source appears variable in near-infrared and possibly in radio as well, on timescales of months to years in the observer's frame. The observed near-infrared spectrum shows no detected emission lines \citep{2022ApJ...929L...7K}. Its redshift was estimated based on the  Gunn--Peterson trough \citep{1965ApJ...142.1633G} found at the wavelength $\lambda = 0.921~\mu\mathrm{m}$, giving a lower redshift limit of $z=6.57$. The observed properties together suggest that this AGN belongs to a subclass of blazars, BL Lac objects, and thus might be the highest-redshift BL Lac discovered to date. Notably, no BL Lac objects have been found so far beyond $z=4$ \citep{2022ApJ...929L...7K}.

VLBI imaging observations with milliarcsec (mas) resolution are essential for verifying the truly compact radio structure and the high brightness temperature expected from a relativistically beamed jet of a BL Lac object. 
Here we report on our dual-frequency ($1.6$ and $4.9$~GHz) observations of J2331$+$1129 with the U.S. National Radio Astronomy Observatory (NRAO) Very Long Baseline Array (VLBA). Throughout this Letter, we assume a flat $\Lambda$ Cold Dark Matter cosmological model with Hubble constant $H_0 = 70$~km\,s$^{-1}$\,Mpc$^{-1}$, matter density parameter $\Omega_\mathrm{m}= 0.3$, and vacuum energy density parameter $\Omega_{\Lambda} = 0.7$. In this model, the luminosity distance at $z=6.57$ is $D_\mathrm{L}= 64.119$\,Gpc, the angular scale $5.425$\,pc\,mas$^{-1}$, and the age of the Universe $813$~Myr \citep{2006PASP..118.1711W}.

\section{Observations and data reduction}
\label{obs}

\begin{figure*}
\centering
\includegraphics[width=0.45\linewidth]{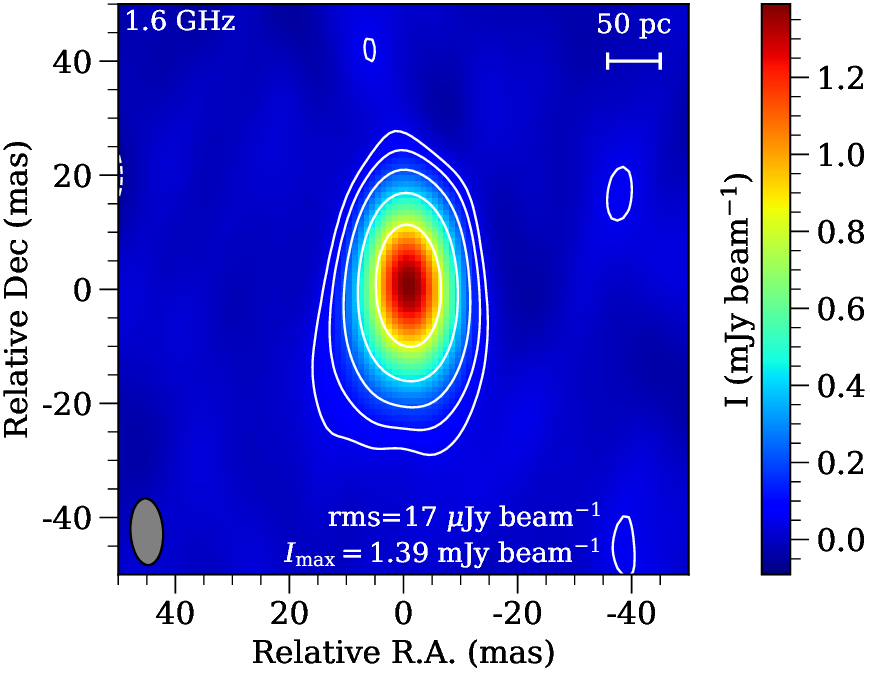}
\includegraphics[width=0.45\linewidth]{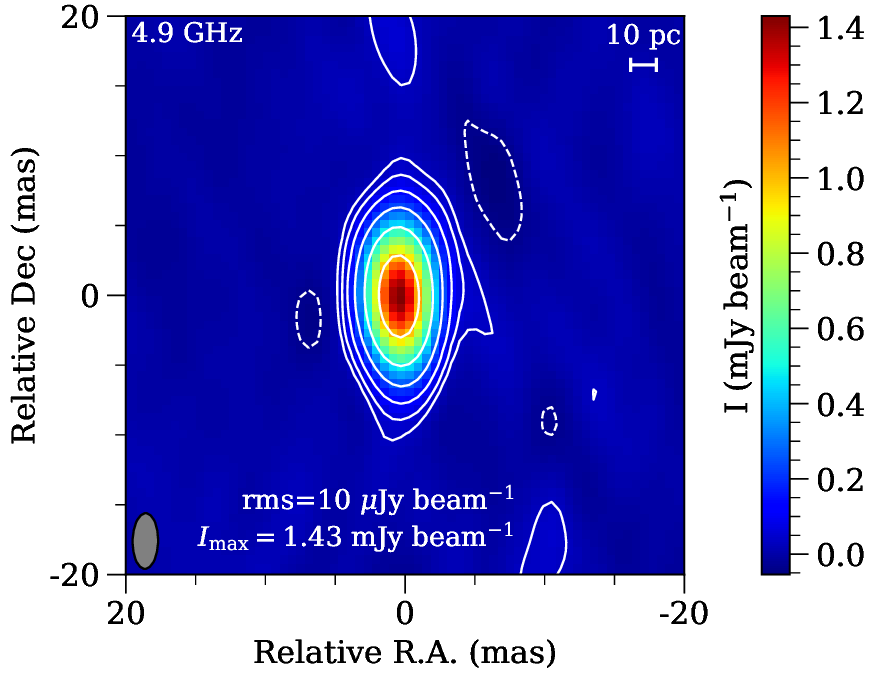}
\caption{Naturally weighted VLBA images of J2331$+$1129 at 1.6~GHz \textit{(left)} and 4.9~GHz \textit{(right)}. The intensity colour scales are displayed on the right-hand side of the panels. The lowest contours start at $\pm3\sigma$ rms noise, and the positive levels increase by a factor of 2. Image rms and peak intensity values are shown in the bottom of the panels, and the elliptical Gaussian restoring beams in the lower left corners. The beam sizes are $11.7~\mathrm{mas} \times 5.7~\mathrm{mas}$ and $4.0~\mathrm{mas} \times 1.8~\mathrm{mas}$ (FWHM), with major axis position angles $3\fdg2$ and $-1\fdg0$ (measured from north through east) at 1.6 and 4.9~GHz, respectively.}\label{fig:image}
\end{figure*}

We conducted observations of J2331$+$1129 with the VLBA at the central frequencies of $1.6$ and $4.9$~GHz on 2022 Feb 1 and 4, respectively (project code: BF132, P.I.: S. Frey). The experiment was performed in phase-referencing mode \citep{1995ASPC...82..327B} using J2330$+$1100 as phase calibrator. This source is listed in the 3rd edition of the International Celestial Reference Frame \citep[ICRF3,][]{2020A&A...644A.159C} and its angular separation from the target is $0\fdg59$. Both experiment segments lasted for $4$~h and recorded data in left and right circular polarizations. The observing time spent on the target, J2331$+$1129, was $165$~min in each segment. For the lower frequency band, the observations were conducted in $2$ intermediate frequency channels (IFs) around $1.4$ and $1.7$~GHz. The data rate was $2$~Gbps. At $4.9$~GHz, the data rate was $4$~Gbps, and the number of IFs $4$. In both experiment segments, the bandwidth was $128$~MHz per IF and polarization. The raw data recorded at each VLBA telescope were correlated at the DiFX correlator \citep{2011PASP..123..275D} in Socorro (New Mexico, USA) with $2$~s integration time. 

The VLBA data were calibrated in the standard way in the NRAO Astronomical Image Processing System package \citep[\textsc{aips},][]{2003ASSL..285..109G}. The appropriate tasks and procedures were used to correct for the dispersive ionospheric delays based on models from Global Navigation Satellite Systems data, and the accurate Earth orientation parameters. After digital sampling corrections, we performed manual phase calibration to remove instrumental delays using a short 1-min scan spent on the secondary calibrator source J2327$+$0940 also scheduled in the experiment. Bandpass correction was performed using the same section of data. Then amplitude calibration followed, using antenna-based system temperatures and gain curves supplied along with the correlated visibility data. Finally, global fringe-fitting was attempted for the bright sources observed, J2330$+$1100, J2327$+$0940, and the fringe-finder 3C\,454.3. The phase, delay and delay-rate solutions were applied to the respective source data. 

The calibrated single-source visibility files were then exported to the Caltech \textsc{Difmap} program \citep{1997ASPC..125...77S} for hybrid mapping. The procedure started with several iterations of \textsc{clean} image decomposition and phase self-calibration. To begin the amplitude and phase self-calibration, the overall antenna gain correction factors were determined. In a few cases when gain corrections exceeded $\pm5\%$, those values were fed back into \textsc{aips} to refine the amplitude calibration.

Fringe-fitting was then repeated for the phase-reference calibrator J2330$+$1100 in \textsc{aips}. Now the \textsc{clean} image of the source obtained in \textsc{Difmap} was also used as an input, to account for a potential small phase contribution of the source structure. The solutions obtained for the calibrator were then interpolated to the target source.

During imaging of J2331$+$1129 in \textsc{Difmap}, we did not attempt any self-calibration because of the weakness of the target source. Radio emission was clearly detected both at $1.6$ and $4.9$~GHz. However, the source position was offset by $\sim 0.1\arcsec$ in both right ascension (RA) and declination (Dec) with respect to the a-priori pointing position. To minimize smearing effects, we repeated the calibration steps in \textsc{aips} but with the target source coordinates shifted to the newly determined accurate position.

\section{Results and discussion}
\label{res}

Results of the \textsc{clean} imaging of J2331$+$1129 are displayed in Fig.~\ref{fig:image}. We found that the source is compact but slightly resolved at both observing frequencies. To characterise its brightness distribution quantitatively, we fitted circular Gaussian model components directly to the interferometric visibility data in \textsc{Difmap}. At $1.6$~GHz, the flux density is $S_\mathrm{1.6\,GHz}=(1.8 \pm 0.2)$~mJy and the component diameter is $\theta_\mathrm{1.6\,GHz} = (3.67 \pm 0.05)$~mas (full width at half maximum, FWHM). At $4.9$~GHz, the parameters of the circular Gaussian component that adequately describe the source are $S_\mathrm{4.9\,GHz}=(1.6 \pm 0.2)$~mJy and $\theta_\mathrm{4.9\,GHz} = (0.68 \pm 0.01)$~mas. The fitted component sizes exceed the minimum resolvable angular size of the interferometer \citep{2005AJ....130.2473K}. These angular sizes correspond to about $(4-20)$~pc linear extent if we assume that the source is at $z=6.57$.

The accurate coordinates of the target source calculated by finding the 4.9-GHz image brightness peak position with the \textsc{aips} verb \textsc{maxfit} are $\mathrm{RA}=23^{\mathrm{h}} 31^{\mathrm{m}} 53\fs21007$ and $\mathrm{Dec}= 11\degr 29\arcmin 52\farcs0089$. The estimated uncertainty of $0.2$~mas in both coordinates is dominated by the effect arising from the angular distance of the phase-reference calibrator source \citep[cf.][]{2004ApJ...604..339C,2020A&ARv..28....6R}. The position determined at $1.6$~GHz is slightly less accurate but the coordinates are consistent with the above values within the errors.

Based on the fitted flux densities, the radio spectrum of the $\sim 10$-pc scale structure is flat ($\alpha_\mathrm{pc}=-0.11$), similar to what is known from the total flux density measurements \citep{2022ApJ...929L...7K}. Irrespective of whether we consider a $\sim 10-20\%$ coherence loss \citep{2010A&A...515A..53M} in phase-referenced observations or not, the measured flux densities match well those obtained with lower-resolution connected-element interferometers, indicating that the total radio emission of J2331$+$1129 is essentially confined to the region imaged with VLBI.  

For comparison with other high-redshift radio AGN, we calculated the $1.4$-GHz rest-frame radio power as $P=4\pi S_\mathrm{1.4\,GHz} D_\mathrm{L}^2(1+z)^{-\alpha-1}$. The high value of $P=1.4\times 10^{26}~$W\,Hz$^{-1}$ suggests that the emission originates from a powerful active galactic nucleus \citep[see][and references therein]{2019MNRAS.490.2542P}, provided that the source is indeed at such a high redshift.

The brightness temperatures were calculated as
\begin{equation}
T_{\mathrm{b}} =1.22 \cdot 10^{12} (1 + z) \frac{S_{\nu}}{\nu^2 \theta^2} \,\text{K,}    
\end{equation}
where the flux density $S_{\nu}$ is measured in Jy, the observing frequency $\nu$ in GHz, and the circular Gaussian component diameter $\theta$ (FWHM) in mas. 
The values ($T_\mathrm{b} \approx 5 \times 10^8$~K at $1.6$~GHz and $T_\mathrm{b} \approx 1.5 \times 10^9$~K at $4.9$~GHz) also confirm the non-thermal nature of the radio emission that originates from AGN activity. This conclusion is so robust that it would hold true even if $z=0$ because brightness temperatures exceeding $T_\mathrm{b} \sim 10^5$~K cannot be attributed to star-forming activity in the host galaxy \citep{1992ARA&A..30..575C}.

Even the higher measured brightness temperature for J2331$+$1129 is an order of magnitude below the equipartition limit, $T_\mathrm{b,eq} \approx 5 \times 10^{10}$~K \citep{1994ApJ...426...51R}, when the energy density in the radiating particles matches that of the magnetic field. This is conventionally assumed as the intrinsic value in the core region of the jet. In our case, the Doppler boosting factor, i.e. the ratio of measured and intrinsic brightness temperatures, $\delta= T_\mathrm{b}/T_\mathrm{b,int} = T_\mathrm{b}/T_\mathrm{b,eq}$, is smaller than $1$. This would indicate that the radio emission is not relativistically boosted, contrary to one would expect from a powerful blazar jet pointing nearly to the line of sight. Even considering somewhat lower intrinsic brightness temperature, $T_\mathrm{b,int} \approx 3 \times 10^{10}$~K \citep{2006ApJ...642L.115H}, found in samples of powerful jetted AGN in their median-low state (i.e. not in outburst when brightness temperatures can reach an order of magnitude higher values), $\delta$ remains well below unity. 

Based on a large sample of jetted AGN (mostly blazars) studied with VLBI, \citet{2020ApJS..247...57C} found that brightness temperatures are generally frequency-dependent, reaching their maximum at the emitted (rest-frame) frequency $\nu_\mathrm{em} \approx 6.8$~GHz. If the redshift of J2331$+$1129 is indeed extremely high, the observed $\nu=4.9~\mathrm{GHz}$ frequency corresponds to $\nu_\mathrm{em} = \nu\,(1+z) = 37~\mathrm{GHz}$ in the expanding Universe. According to the phenomenological dependence derived by \citet{2020ApJS..247...57C}, the maximum brightness temperature of the source could then be a factor of $\sim 4-5$ higher than what we measured at $\nu=4.9$~GHz, i.e. $\sim5 \times 10^9$~K. In any case, this is still not sufficient for having observed Doppler-boosted emission in the J2331$+$1129 jet. We note, however, that the core brightness temperatures (or lower limits) of the only other claimed blazar candidate known to date at $z>6$, PSO~J030947.49$+$271757.31, are also in the order of $T_\mathrm{b} \sim 10^8$~K \citep{2020A&A...643L..12S}. On the other hand, for the previously known second most distant blazar, J0906$+$6930 $(z=5.47)$, the brightness temperature measured with VLBI ($3 \times 10^{11}$~K) clearly exceeds the equipartition limit \citep{2020NatCo..11..143A}. 

The flat section of the radio spectrum of J2331$+$1129 around $\nu \ga 1$~GHz \citep[see also][]{2022ApJ...929L...7K}, although not yet sampled in a sufficiently broad range of frequencies, could possibly be interpreted as similar to those of Gigahertz-Peaked Spectrum (GPS) radio sources \citep[e.g.][and references therein]{2021A&ARv..29....3O}. However, GPS sources typically have more extended mas-scale radio structure, within about $500$~pc in size \citep{2021A&ARv..29....3O}. Indeed, a high-redshift GPS source recently studied with VLBI, J1606$+$3124 $(z=4.56)$, shows a relatively compact ($\sim 70$~pc) triple structure \citep{2022MNRAS.511.4572A}, much different from the single component ($<20$~pc) we see in J2331$+$1129.

Note that, assuming $z=6.57$ for our source, the observed $\sim 1$~GHz frequency range translates to above $5$~GHz, where turnover frequencies are characteristic to High-Frequency Peaker (HFP) sources \citep{2021A&ARv..29....3O}. Typically both HFPs and GPS sources have less compact structures that blazar core--jets. However, HFP and blazar properties are not necessarily mutually exclusive. A good example is J0906$+$6930 $(z=5.47)$ which has a peaked broad-band radio spectrum with a turnover frequency $\nu_{\mathrm{em}} \approx 40$~GHz \citep{2017MNRAS.467.2039C,2017MNRAS.468...69Z} but also very compact Doppler-boosted jet emission \citep{2020NatCo..11..143A}. Future flux density measurements of J2331$+$1129 in a frequency range wider than available at present \citep[like in the case of a sample of $z>5$ radio AGN by][]{2022A&A...659A.159S} would provide decisive information about the possible HFP nature of this source. At present, the previous low-resolution and our recent VLBI data indicate that the radio spectrum of J2331$+$1129 is likely flat between  the observed $888$~MHz and $4.9$~GHz, as expected for blazars. The currently available data do not show evidence for a spectral turnover at gigahertz frequencies.

\section{Summary and conclusions}
\label{concl}

The redshift of $z=6.57$ for the radio-emitting AGN J2331$+$1129 was inferred from the shape of the near-infrared continuum spectrum \citep{2022ApJ...929L...7K}. The lack of identifiable emission lines, as well as the flat radio spectrum in the observed GHz range, the spectral energy distribution, and the indication of variability led \citet{2022ApJ...929L...7K} to classify the source as a candidate BL Lac object, by far the most distant known to date. However, since the redshift determination is based on locating the break in the continuum spectrum and not emission lines, the value should be considered approximate.  

We performed high-resolution radio interferometric observations with the VLBA to reveal whether there is a compact core--jet structure in J2331$+$1129. Nearly simultaneous dual-frequency ($1.6$- and $4.9$-GHz) imaging showed a compact, flat-spectrum radio core whose flux density matches the total flux density, indicating that the entire radio emission is confined within $\sim20$~pc. Via traditional nodding-style phase-referencing to a nearby ICRF3 radio quasar, we precisely determined the astrometric position of the source with sub-mas accuracy. 

The inferred maximum brightness temperature is about $5 \times 10^9$~K, confirming that this is an AGN source but providing no clear evidence for Doppler-boosted radio jet emission. This leaves the BL Lac identification of the source an open question. Comparing the brightness temperature of J2331$+$1129 with those of other high-redshift ($z \ga 6$) blazars would be critical for understanding this. However, currently the sample of known sources with similar properties is very small. It is possible that the observed jet properties in extremely distant radio sources differ from those at lower redshifts, due to yet poorly understood evolutionary and environmental effects. The rarity of $z \ga 6$ blazars makes each new discovery and study valuable for our understanding of the behaviour of jetted SMBHs in the early Universe. Observational limitations could also play a role, as phase-referenced VLBI observations of weak radio sources may only provide an upper limit of the resolved source size, and thus a lower limit to the brightness temperature.

To verify the BL Lac nature of J2331$+$1129, sensitive VLBI observations at higher frequencies and thus higher resolution might prove useful. Also, densely time-sampled total flux density monitoring could possibly reveal rapid variations indicative of high brightness temperature. Finally, it cannot be excluded that the source is a misidentified BL Lac object and/or it is not at extremely high redshift. This should be tested in the future with sensitive spectroscopic observations, to look for emission lines and possibly re-evaluate the redshift.

\begin{acknowledgements}
 The National Radio Astronomy Observatory is a facility of the National Science Foundation operated under cooperative agreement by Associated Universities, Inc.
 This work made use of the Swinburne University of Technology software correlator \citep{2011PASP..123..275D}, developed as part of the Australian Major National Research Facilities Programme and operated under licence.
 This work was supported by the Hungarian National Research, Development and Innovation Office (OTKA K134213 \& PD146947). This project has received funding from the HUN-REN Hungarian Research Network. T.A. thanks for the financial support from the Pinghu Laboratory. E.K. thanks for support from Ministry of Science and Technology of Taiwan grants MOST 109-2112-M-008-021-MY3 and 112-2112-M-008-017-MY3. Y.Z. is supported by the National SKA Programme of China (grant no. 2022SKA0120102), Shanghai Sailing Program (grant no. 22YF1456100).
\end{acknowledgements}

\bibliographystyle{aa} 
\bibliography{bf132} 

\end{document}